\documentclass[aps,twocolumn,superscriptaddress,showpacs,showkeys]{revtex4}
\usepackage{graphics,graphicx,dcolumn,bm,fleqn,epic,eepic,float}
\usepackage{amssymb,amsmath,multirow,rotate,color,float,times}
\usepackage{color}
\usepackage{soul}             
\usepackage[latin1]{inputenc} 
\definecolor{red}{rgb}{1,0,0}
\definecolor{green}{rgb}{0,1,0}
\definecolor{blue}{rgb}{0,0,1}

\newcommand{\bite}{\begin{itemize}}
\newcommand{\eite}{\end{itemize}}
\newcommand{\benu}{\begin{enumerate}}
\newcommand{\eenu}{\end{enumerate}}
\newcommand{\bverb}{\begin{verbatim}}
\newcommand{\everb}{\end{verbatim}}

\newcommand{\beq}{\begin{equation}}
\newcommand{\eeq}{\end{equation}}
\newcommand{\barr}{\begin{array}}
\newcommand{\earr}{\end{array}}

\newcommand{\f}{\mathbf}

\begin{document}

\title{Principal axes for stochastic dynamics}

\author{V.V.~Vasconcelos}
\affiliation{Physics Department, Faculty of Sciences,
             University of Lisbon, 1649-003 Lisbon, Portugal} 
\author{F.~Raischel}
\affiliation{Center for Theoretical and Computational Physics, 
             University of Lisbon,
             Av.~Prof.~Gama Pinto 2, 1649-003 Lisbon, Portugal}
\author{M.~Haase}
\affiliation{Institute for High Performance Computing, University of Stuttgart,
             Nobelstr.~19, DE-70569 Stuttgart, Germany}
\author{J.~Peinke}
\affiliation{Institute of Physics, University of Oldenburg,
             DE-26111 Oldenburg, Germany}
\author{M.~W\"achter}
\affiliation{Institute of Physics, University of Oldenburg,
             DE-26111 Oldenburg, Germany}
\author{P.G.~Lind}
\affiliation{Physics Department, Faculty of Sciences,
             University of Lisbon, 1649-003 Lisbon, Portugal} 
\affiliation{Center for Theoretical and Computational Physics, 
             University of Lisbon,
             Av.~Prof.~Gama Pinto 2, 1649-003 Lisbon, Portugal}
\author{D.~Kleinhans}
\affiliation{Institute for Marine Ecology, University of Gothenburg,
             Box 461, SE-405 30 G\"oteborg, Sweden}
\affiliation{Institute of Theoretical Physics, University of
             M\"unster, DE-48149 M\"unster, Germany}

\begin{abstract}
We introduce a general procedure for directly ascertaining how many independent
stochastic sources exist in a complex system modeled through a set of coupled 
Langevin equations of arbitrary dimension. 
The procedure is based on the computation of the eigenvalues and the
corresponding eigenvectors of local diffusion matrices. We demonstrate our 
algorithm by applying it to  two examples of systems showing Hopf-bifurcation.
We argue that computing the eigenvectors associated to the eigenvalues
of the diffusion matrix at local mesh points in the phase space
enables one to define vector fields of stochastic eigendirections.
In particular, the eigenvector associated to the lowest eigenvalue defines
the path of minimum stochastic forcing in phase space, and a transform to a 
new coordinate system aligned with the eigenvectors can increase the 
predictability of the system.
\end{abstract}



\maketitle


\section{Introduction}
\label{sec:int}

When dealing with measurements on complex systems it is typically difficult 
to find the optimal set of variables to describe their evolution, which 
provides the best opportunities for understanding and predicting the system's 
behavior.

Recently a framework for analyzing measurements
on stochastic systems was introduced \cite{friedrich97,friedrich08}. 
This framework is based on 
the assumption that the measured properties evolve according to a
deterministic drift and stochastic fluctuations due to the interactions
with the internal degrees of freedom or the environment.
These fluctuations are globally ruled by some specific stochastic 
forcing inherent to the signal evolution itself, and therefore cannot 
be separated from the measured variables. Whether they are due to quantum uncertainties or a limited knowledge of the state of the system, they are fundamental for a complete description of a complex system.
Several works in this scope have already shown the advantage of 
this approach, ranging from the description of turbulent 
flows \cite{friedrich97} and climate indices \cite{lind05} to the evolution
of stock markets \cite{friedrich00} and oil prices \cite{ghasemi07},
just to mention some. 
In the course of time, 
several improvements to the method were proposed concerning its robustness 
with respect to finite sampling effects and measurement noise
\cite{boettcher06,lind10,carvalho2010,Kleinhans05,Gottschall08NJP,Lade09}. 

However, a natural question arises when dealing simultaneously with several
variables: is it 
possible to find non-trivial functions of measured properties for which the stochastic fluctuation 
can be neglected? This would mean that it should be possible to decrease
the number of stochastic variables needed to describe the system.

In this paper we will follow this question up in detail and show, that
by accessing the eigenvalues of the diffusion matrices comprehending the
measured properties it is possible to derive a path in phase space through 
which the deterministic contribution is enhanced. As a direct application,
our procedure allows for the determination of the number of independent sources
of stochastic forcing in a system described by an arbitrarily large number
of properties.

A pictorial example is as follows.
Consider a professional archer
trying to aim at a target in a succession of shots.
If the archer holds the bow  without any support, one expects that the
set of trials is distributed around the center according to a radially 
symmetric Gaussian.
Hence deviations have the same amplitudes in all directions.
However, if the archer lies on the floor, the vertical direction will be 
more confined than the horizontal. 
In this case we expect a smaller variance of deviations in the vertical 
direction than in the horizontal one.
If the archer lies on an inclined plane the most confined and less confined
directions will have a certain slope related to the plane inclination.
Regarding the variance in space as a particular 
representation of the diffusion matrix, 
we can now think in more
complex systems where at each point in phase space fluctuations can be defined
through their local diffusion matrices. The study of its eigenvalues and 
eigenvectors is the scope of the present paper.

We start in Sec.~\ref{sec:model} by defining our system of
variables mathematically  and by describing the standard approach to estimate 
its dynamics from measured data as introduced in 
Ref.~\cite{friedrich97,friedrich08}, where a particular emphasis will be 
given to the diffusive terms. 
Then the general eigenvalue problem is introduced in the role of the local 
eigenvalues of diffusion matrices
, which exhibits the framework 
for the present work.
We show that at each point of phase space the eigenvectors of the
diffusion matrix are tangent to new coordinate lines, one of them
corresponding to the lowest eigenvalue, thus indicating the direction 
towards which fluctuations are - at least partially - suppressed.
In Sec.~\ref{sec:hopf} we demonstrate our approach on two examples from 
Hopf-bifurcation systems with stochastic forcing.
Section \ref{sec:conclusions} closes the paper with discussion and
conclusions.

\section{Defining stochastic eigendirections}
\label{sec:model}

We consider an $N$-dimensional Langevin process 
$\mathbf{X}=(X_1(t),\dots,X_N(t))$ whose probability density functions
(PDFs) $f(\mathbf{X},t)$ evolve according to the Fokker-Planck equation 
(FPE) \cite{fpeq,gard} 
\begin{eqnarray}
\frac{\partial f(\mathbf{X},t)}{\partial t} &=&
       -\sum_{i=1}^N\frac{\partial}{\partial x_i}
        \left [ 
           D_i^{(1)}(\mathbf{X})f(\mathbf{X,t})
        \right ] \cr
    & &+\sum_{i=1}^N\sum_{j=1}^N
        \frac{\partial ^2}{\partial x_i\partial x_j}
        \left [
           D_{ij}^{(2)}(\mathbf{X})f(\mathbf{X},t)
        \right ] \quad .
\label{FPE}
\end{eqnarray}
The functions $D_i^{(1)}$ and $D_{ij}^{(2)}$ are called the Kramers-Moyal or 
the drift and diffusion coefficients and are defined as
\begin{equation}
\mathbf{D}^{(k)}(\mathbf{X})=\lim_{\Delta t\rightarrow0}\frac{1}{\Delta t}
                                 \frac{\mathbf{M}^{(k)}(\mathbf{X},\Delta t)}{k!}
\label{DefCoefKM}\quad,
\end{equation}
where $\mathbf{M}^{(k)}$ 
are the first and second conditional moments ($k=1,2$).
Here we assume that the underlying process is stationary and therefore
both drift and diffusion coefficients do not explicitly depend on
time $t$.
Conditional moments
can be directly derived from the measured data as \cite{friedrich08,lind10}:
\begin{equation}
\begin{array}{lcl}
M_i^{(1)}(\mathbf{X},\Delta t) = 
     \left\langle Y_i(t+\Delta t)-Y_i(t)  | {\mathbf{Y}(t)=\mathbf{X}} \right\rangle  & & \\
M_{ij}^{(2)}(\mathbf{X},\Delta t) = & &  \\
     \left\langle (Y_i(t+\Delta t)-Y_i(t))(Y_j(t+\Delta t)-Y_j(t))
     | {\mathbf{Y}(t)=\mathbf{X}}\right\rangle  & & 
\end{array}
\label{2D}
\end{equation}
Here $\mathbf{Y}(t)=(Y_1(t),\dots,Y_N(t))$ exhibits the $N$-dimensional vector of measured variables at the $t$ and $\langle \cdot
| {\mathbf{Y}(t)=\mathbf{X}} \rangle$ symbolizes a conditional averaging over the entire measurement period, where only measurements with ${\mathbf{Y}(t)=\mathbf{X}}$ are taken into account\footnote{In practice binning or kernel based approaches with a certain threshold are applied in order to evaluate the condition $\mathbf{Y}(t)=\mathbf{X}$. See e.g.~Ref.~\cite{Lamouroux09} for details.}.
$\mathbf{D}^{(1)}$ is the drift vector and $\mathbf{D}^{(2)}$ 
the diffusion matrix.
Notice that $\mathbf{Y}(t)=\mathbf{X}$ is considered to be true in a vicinity of $\mathbf{X}$ due to the observational limitations. This vicinity must be small and such that the conditional moments do not change abruptly for small 
$\Delta t$, making $\mathbf{D}^{(1)}$ and $\mathbf{D}^{(2)}$ continuous.

Associated with the FPE (\ref{FPE}) is a system of $N$ coupled 
It\^o-Langevin equations, which can be written as \cite{fpeq,gard} 
\begin{equation}
\frac{d \mathbf{X}}{d t}= 
                     \mathbf{h}(\mathbf{X})
                     + \mathbf{G}(\mathbf{X}) 
                     \mathbf{\Gamma}(t)\quad.
\label{Lang2DVect}
\end{equation}
Here $\mathbf{\Gamma}(t)$ is a set of $N$ normally distributed random variables fulfilling 
\begin{equation}
\langle \Gamma_i(t)\rangle=0, 
\quad \langle \Gamma_i(t)\Gamma_j(t')\rangle=2\delta_{ij}\delta(t-t')\quad,
\label{gamma}
\end{equation}
that drives the stochastic evolution of $\mathbf{X}$. The vectors $\mathbf{h}$ and the matrices $\mathbf{G}=\{ g_{ij} \}$ for all $i,j=1,\dots,N$
are connected to the local drift and diffusion function through 
\begin{eqnarray}
\label{Deqh1}
D_i^{(1)}(\mathbf{X}) &=& h_i (\mathbf{X})\quad\mbox{and}\\
D^{(2)}_{ij}({\bf X})&=&\sum^N_{k=1}g_{ik}({\bf X})g_{jk}({\bf X}) \quad.
\label{DeqG2}
\end{eqnarray}

Such methodology is applicable under certain conditions, namely the
markovian nature of the underlying process. Further, the
FPE is only valid as long as $D^{(k)}\sim 0$ for $k\geq3$. 
In case
the stochastic force has zero average and is Gaussian and $\delta$-correlated
a Langevin differential equation is obtained. In case the stochastic
forces $\mathbf{\Gamma}$ not Gaussian $\delta$-correlated, obtaining 
a stochastic equation is still possible but with stochastic forcing
characterized by a L\'evy-stable distribution\cite{friedrich08,lind10}.
While the FPE describes the evolution of the joint distribution of the $N$ 
variables statistically, the system of Langevin equations in 
(\ref{Lang2DVect}) models individual stochastic trajectories of the system.
In Eq.~(\ref{Lang2DVect}) the term $\mathbf{h(X)}$ contains the
deterministic part of the macroscopic dynamics,
while the 
functions $\mathbf{G(X)}$ account for the amplitudes 
of the stochastic forces mirroring the different sources 
of fluctuations due to all sorts of microscopic interactions within the 
system. 
Here we should point out that we consider stationary processes, otherwise 
ensemble averages have to be taken\cite{friedrich08}.
Since $D^{(1)}$  and $D^{(2)}$ are considered to be continuous,
the conditional moments should be differentiable
(see Eq.~(\ref{DefCoefKM})), which means
that in a sufficiently small time interval the dependence of the conditional
moments on the time increment is linear.
\begin{figure}[b]
\begin{center}
\includegraphics*[width=0.49\textwidth,angle=0]{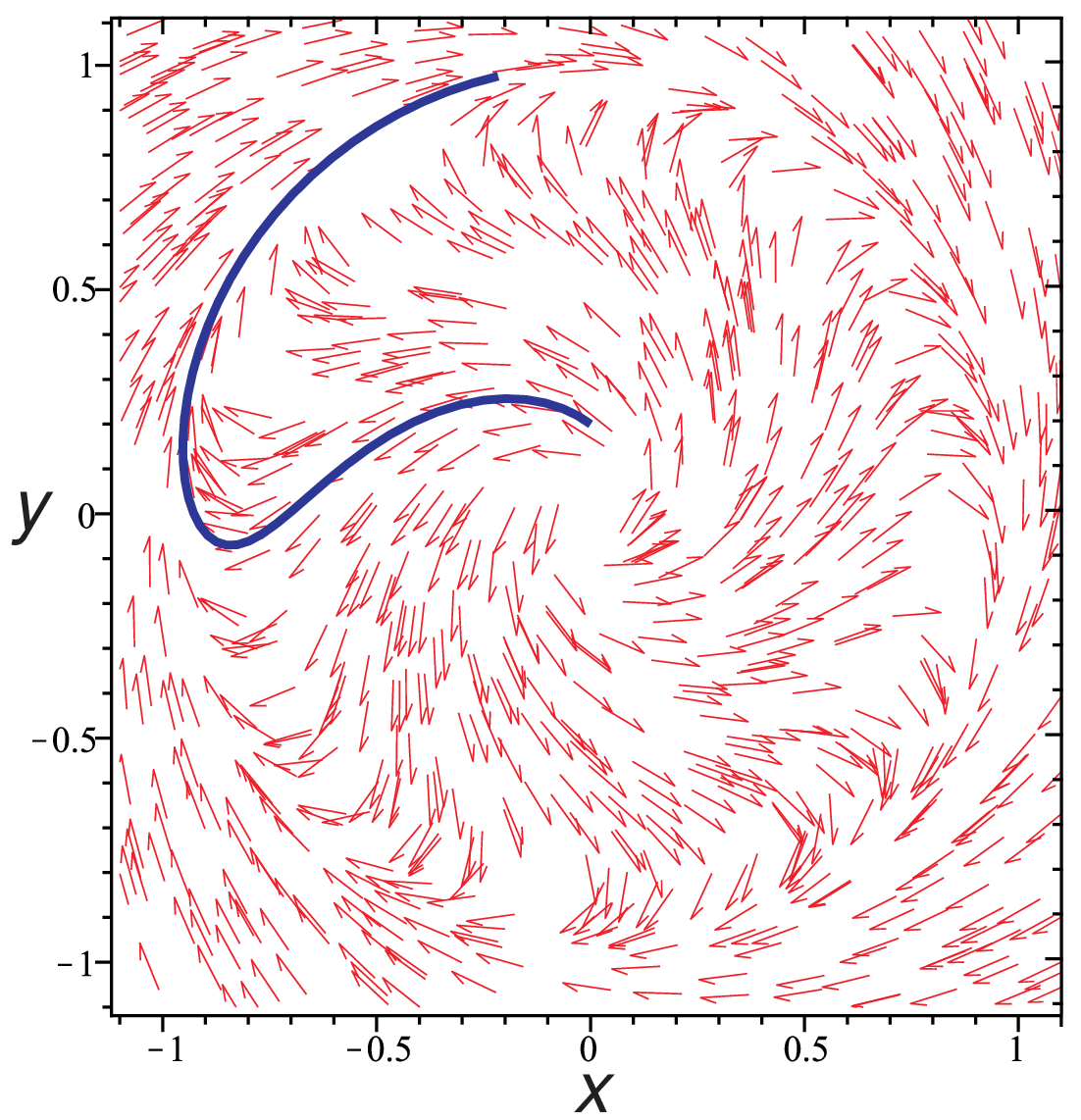}
\end{center}
\caption{\protect
         (Color online)
         Stream plot of the system in Eq.~(\ref{HopfSys}) with no 
         stochasticity ($\mathbf{D}^{(2)}\equiv 0$), i.e.~$k_1=k_2=0$, where the 
         arrows formally represent the drift vector of synthetic data 
         extracted from direct simulation of Eq.~(\ref{HopfSys}).
         The solid line represents a partial trajectory.
         Trajectories diverge from the unstable fixed point
         $r=0$ and converge to the limit cycle $r=1$.}
\label{fig1}
\end{figure}
\begin{figure}[htb]
\begin{center}
\includegraphics*[width=0.48\textwidth,angle=0]{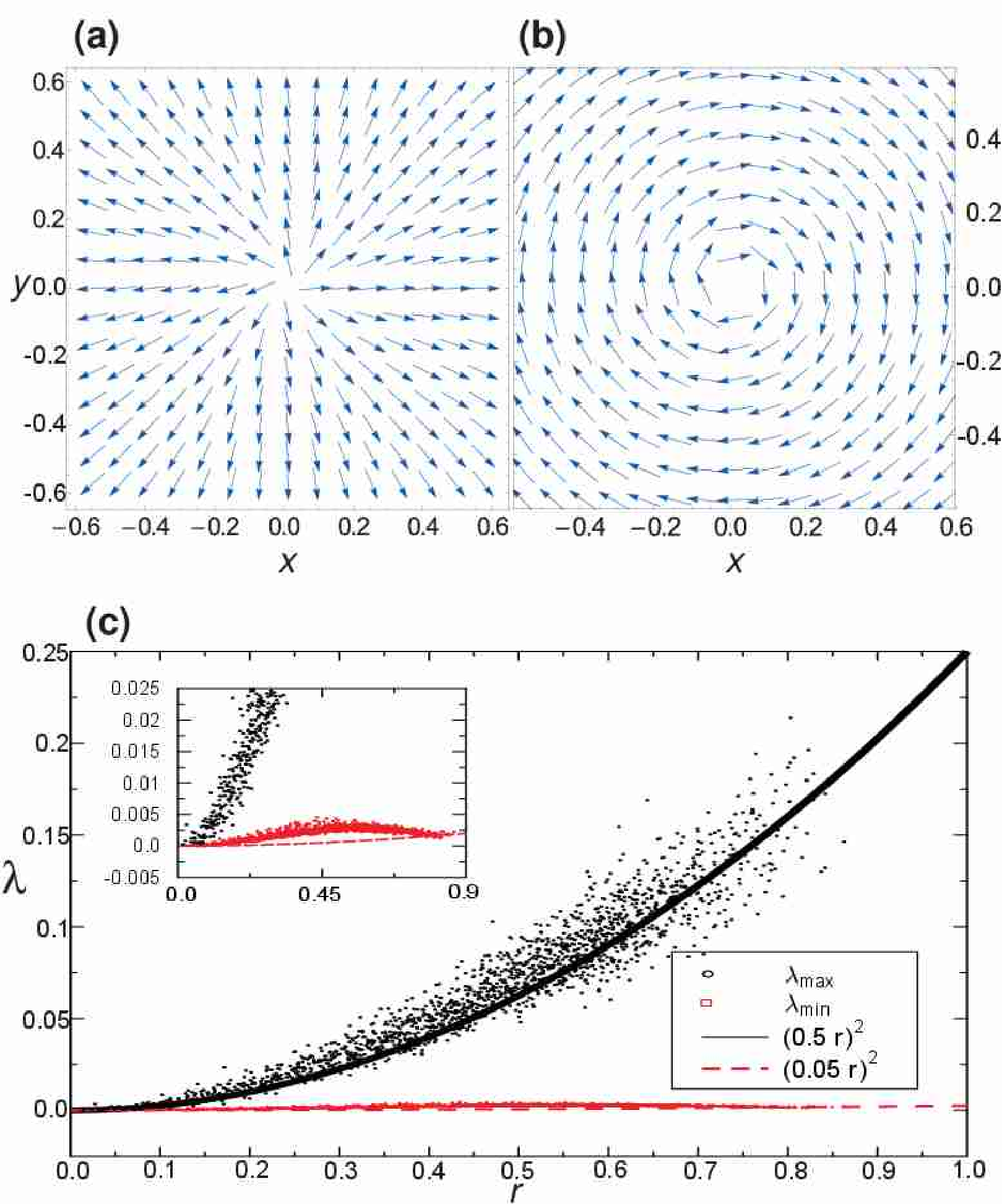}
\end{center}
\caption{\protect
         (Color online)
         Principal axis of the diffusion matrix.
         Field of 
         {\bf (a)} largest and 
         {\bf (b)} smallest 
         eigenvectors for the diffusion matrix corresponding 
         to  Eq.~(\ref{HopfSys}) together with
         {\bf (c)} the eigenvalues $\lambda$ as a function of $r$
         and (inset) a close-up around $\lambda\sim 0$ (see text).
         The largest eigenvalue is associated with the radial direction
         while the smallest one corresponds to the angular one.
         Here $k_1=0.5$, $k_2=0.05$,
         $\alpha = 0.7475 \approx \left\langle r^2\right\rangle$,
         $t_0=0$, $r_0=1.01$ and $\theta_0=0$. The analysis was
         performed from $10^7$ data points extracted by integration
         of Eq.~(\ref{HopfSys}) with integration step of 
         $\Delta t=10^{-4}$.}
\label{fig2}
\end{figure}

The conditional moments in Eq.~(\ref{2D}) are computed directly from data time series and are typically linear functions of $\Delta t$ for sufficiently
small $\Delta t$ \cite{friedrich08,lind10}. Then, through the limit in
Eq.~(\ref{DefCoefKM}), both $D^{(1)}_i$ and $D^{(2)}_{ij}$ are determined. 
When dealing with multidimensional systems which have associated, 
at least \textit{a priori}, several independent sources of stochastic
fluctuations, the observation time may need to be considerably large,
to sample in sufficient detail the system dynamics. Typically,
data sets should be not only large but also stationary. Though, recently
new developments to such methods were done to suit more general situations
were time series are short and non-stationary\cite{mourik06}.

The $N \times N$ matrix $\mathbf{G}$ cannot be uniquely
determined from the symmetric diffusion matrix $\mathbf{D^{(2)}}$ for 
$N \ge 2$, because the number of unknown elements in $\mathbf{G}$ exceeds 
the number of known elements in $\mathbf{D^{(2)}}$ leading to  
$N^2-\frac{1}{2}N(N+1)=\frac{1}{2}N(N-1)$ free parameters. 
However, a simple method to obtain $\mathbf{G}$ from $\mathbf{D^{(2)}}$ is 
the following. 
Due to its symmetry and positive semi-definiteness the diffusion
matrix  $\mathbf{D^{(2)}}$ has only real, non-negative eigenvalues 
$\lambda_i (i=1,\ldots,N)$.
Therefore an orthogonal transformation $\mathbf{U}$ can be found that 
diagonalizes $\mathbf{D^{(2)}}$, 
i.e.~$\mathbf{U}^T\mathbf{D^{(2)}}\mathbf{U}=\rm{diag(\lambda_1,\ldots,\lambda_N)}$.
Taking the positive root of the eigenvalues and transforming it back we
arrive at $g_{ij}=(\sqrt{\mathbf{D^{(2)}}})_{ij}$ where the symbolic notation
$\sqrt{\mathbf{D^{(2)}}}=\mathbf{U}{\rm{diag(\sqrt{\lambda_1},\ldots,
\sqrt{\lambda_N})}}\mathbf{U}^T$ is used for simplicity.
General forms of $\mathbf{G}$ can be constructed by multiplication of 
$\sqrt{\mathbf{D^{(2)}}}$ with arbitrary orthogonal matrices. 
For our considerations the simple version is sufficient
\cite{fpeq}.

Since $\mathbf{h}$ and $\mathbf{G}$ are functions of the 
variables $\mathbf{X}$ and are numerically determined on a 
$n_1\times...\times n_N$ 
mesh of points in phase space, one can always 
define at each mesh point the $N$ eigenvalues and corresponding 
eigenvectors of the matrix $\mathbf{G} (\mathbf{X})$.
This analysis provides information about the stochastic forcing acting on
the system and was already applied to a two-dimensional sub-critical 
bifurcation \cite{gradisek_eigenvectors} and to the analysis of human 
movements \cite{vanMourik_eigenvectors}.

In general
the eigenvalues indicate the amplitude of the stochastic force and the
corresponding eigenvector indicates the direction toward which 
such force acts.
Even more interesting features, however, can be extracted from the 
eigenvalues and eigenvectors.

To each eigenvector of the diffusion matrix we can
associate one independent source of stochastic forcing $\Gamma_i$.
In this scope, the eigenvectors can be regarded as defining
principal axes for stochastic dynamics.
For instance, the vector field aligned at each mesh point to the
eigenvector associated to the smallest eigenvalue of matrix $\mathbf{G}$ 
defines the paths in phase space towards which the fluctuations are
minimal.
Furthermore, if the corresponding eigenvalues are very small compared 
to all the other ones at the respective mesh points, 
the corresponding stochastic forces can be neglected and the system 
has only $N-1$ independent stochastic forces. 
In this case the problem
can be reduced in one stochastic variable by an appropriate transformation of variables, since the eigenvectors 
in one coordinate system are the same as in another one 
(see Append.~\ref{app:coordinate}).

These are the central ideas of our study which we next apply to an
analytical example, namely the Hopf-bifurcation.

\section{The Hopf-bifurcation system with stochastic forcing: 
         an analytical example}
\label{sec:hopf}

In what follows we consider the dynamical system
\begin{subequations}
\begin{eqnarray}
\frac{dr}{dt}&=&r(1-r^2) + k_{1}r \Gamma_1 \\
\frac{d \theta}{dt}&=&\alpha -r^2 + k_{2} \Gamma_2 \label{thetaeq}
\end{eqnarray}
\label{HopfSys}
\end{subequations}
describing the evolution of the radial and azimuthal coordinates of a 
particle moving in two-dimensional space, with $\alpha$, $k_1$ and $k_2$
being constants. $\Gamma_{1,2}$ are the stochastic forces, which are 
Gaussian distributed and $\delta$-correlated.
For $k_{1}=k_{2}=0$ the system for $\alpha>0$ has
an unstable fixed point at $r=0$ and 
a stable limit cycle at $r=1$, 
as sketched in Fig.~\ref{fig1}. Equation (\ref{thetaeq}) uses 
$\alpha-r^2$ instead of the usual $1+r^2$ \cite{mariabook} in order to be able to avoid the 
systematic increase of $\theta$ in time. For this reason we chose $\alpha = \langle r^2 \rangle$.

\begin{figure}[t]
\begin{center}
\includegraphics*[width=0.48\textwidth,angle=0]{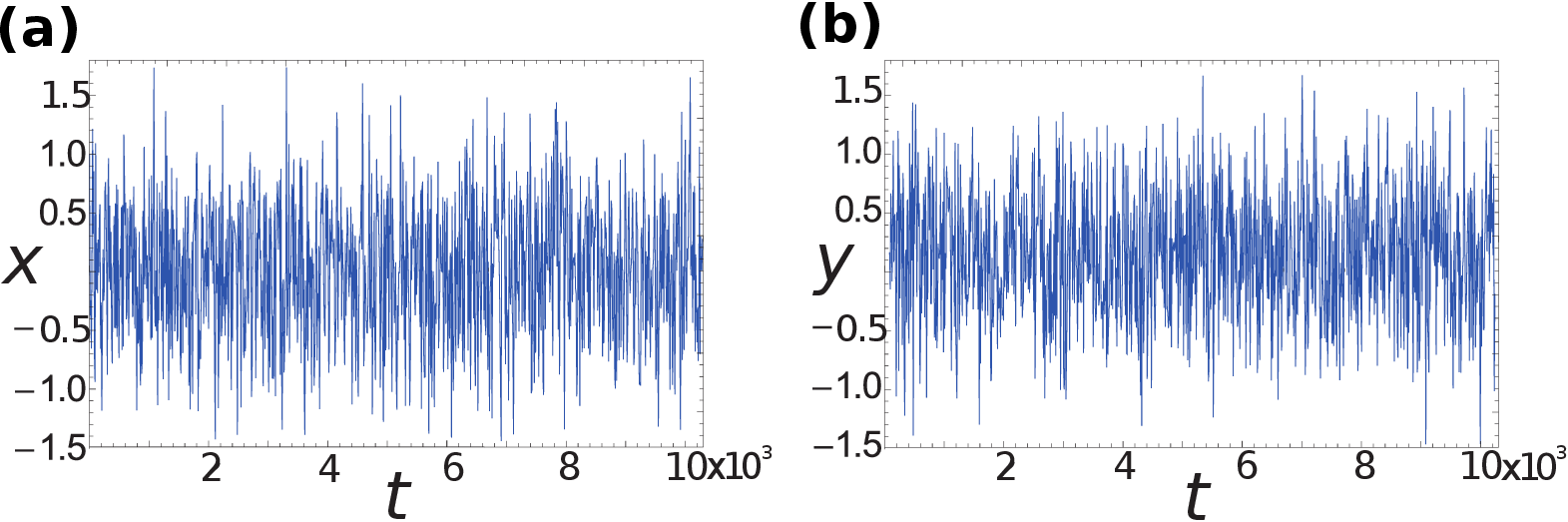}
\end{center}
\caption{\protect
         (Color online)
         Time series of 
         {\bf (a)} $x(t)=r(t)\cos{\theta(t)}$ and 
         {\bf (b)} $y(t)=r(t)\sin{\theta(t)}$
         where $r$ and $\theta$ are taken from the integration of 
         Eq.~(\ref{HopfSys}) for the same conditions as in 
         Fig.~\ref{fig2}.}
\label{fig3}
\end{figure}
\begin{figure*}   
\begin{center} 
\includegraphics*[width=\textwidth,angle=0]{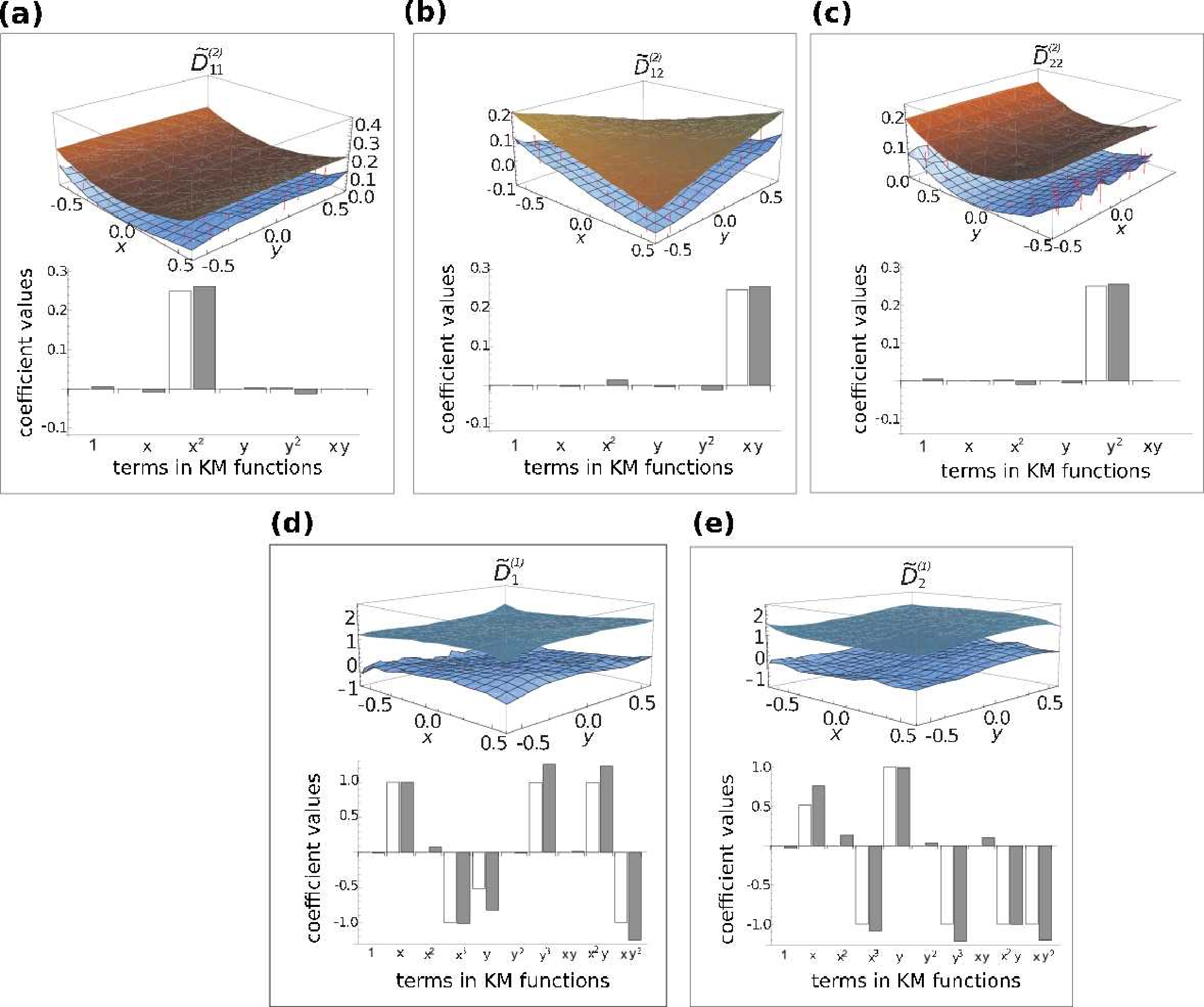}
\end{center}
\caption{\protect  
         (Color online)
         Drift and diffusion functions for the system 
         in Eq.~(\ref{HopfSys}) for the same conditions as in Fig.~\ref{fig2},
         obtained from analysis of the time series $x=r\cos{\theta}$ and $y=r\sin{\theta}$.
         For each drift and diffusion function the bottom surface
         reflect the numerical results, while the corresponding fitted surface 
         on top is shifted upwards for clarity:
         {\bf (a)} $\tilde{D}_{11}^{(2)}(x,y)$,
         {\bf (b)} $\tilde{D}_{12}^{(2)}(x,y)$,
         {\bf (c)} $\tilde{D}_{22}^{(2)}(x,y)$,
         {\bf (d)} $\tilde{D}_{1}^{(1)}(x,y)$,
         {\bf (e)} $\tilde{D}_{2}^{(1)}(x,y)$.
         The bar plots compare the coefficients that fit each function 
         (dark bars) with the analytic, true ones (white bars) (see Tab.~\ref{tab1}).}
\label{fig4}
\end{figure*}
\begin{table*}[htb]
\begin{tabular}{|c||c|c|c|c|c|c|c|c|c|c|}
\hline
 & \multicolumn{10}{c|}{\ } 
\\
\multicolumn{1}{|c||}{Kramer-Moyal} &
\multicolumn{10}{c|}{Coefficients for} 
\\
\multicolumn{1}{|c||}{functions} & 
\multicolumn{10}{c|}{\ } 
\\\cline{2-11}
\multicolumn{1}{|c||}{\ } & 
\multicolumn{1}{c|}{$1$} & 
\multicolumn{1}{c|}{$x$} & 
\multicolumn{1}{c|}{$x^2$} & 
\multicolumn{1}{c|}{$x^3$} & 
\multicolumn{1}{c|}{$y$} & 
\multicolumn{1}{c|}{$y^2$} & 
\multicolumn{1}{c|}{$y^3$} & 
\multicolumn{1}{c|}{$xy$} & 
\multicolumn{1}{c|}{$x^2y$} & 
\multicolumn{1}{c|}{$xy^2$} 
\\\hline
\multicolumn{1}{|c||}{$\tilde{D}^{(1)}_1$} & 
\multicolumn{1}{c|}{$-0.009$} &
\multicolumn{1}{c|}{$1.00$} &
\multicolumn{1}{c|}{$0.077$} & 
\multicolumn{1}{c|}{$-1.01$} & 
\multicolumn{1}{c|}{$0.820$} & 
\multicolumn{1}{c|}{$-0.012$} &
\multicolumn{1}{c|}{$1.26$} &
\multicolumn{1}{c|}{$0.014$} &
\multicolumn{1}{c|}{$1.23$} &
\multicolumn{1}{c|}{$1.24$} 
\\\hline
\multicolumn{1}{|c||}{$\tilde{D}^{(1)}_2$} & 
\multicolumn{1}{c|}{$-0.034$} &
\multicolumn{1}{c|}{$0.761$} &
\multicolumn{1}{c|}{$0.133$} & 
\multicolumn{1}{c|}{$-1.09$} & 
\multicolumn{1}{c|}{$0.99$} & 
\multicolumn{1}{c|}{$0.036$} &
\multicolumn{1}{c|}{$-1.22$} &
\multicolumn{1}{c|}{$0.101$} &
\multicolumn{1}{c|}{$-1.00$} &
\multicolumn{1}{c|}{$-1.21$} 
\\\hline
\multicolumn{1}{|c||}{$\tilde{D}^{(2)}_{11}$} & 
\multicolumn{1}{c|}{$0.005$} &
\multicolumn{1}{c|}{$-0.008$} &
\multicolumn{1}{c|}{$0.262$} & 
\multicolumn{1}{c|}{$\sim 0$} &
\multicolumn{1}{c|}{$0.001$} & 
\multicolumn{1}{c|}{$0.013$} &
\multicolumn{1}{c|}{$\sim 0$} &
\multicolumn{1}{c|}{$\sim 0$} &
\multicolumn{1}{c|}{$\sim 0$} &
\multicolumn{1}{c|}{$\sim 0$} 
\\\hline
\multicolumn{1}{|c||}{$\tilde{D}^{(2)}_{12}=\tilde{D}^{(2)}_{21}$} & 
\multicolumn{1}{c|}{$\sim 0$} &
\multicolumn{1}{c|}{$-0.003$} &
\multicolumn{1}{c|}{$014$} & 
\multicolumn{1}{c|}{$\sim 0$} &
\multicolumn{1}{c|}{$-0.004$} & 
\multicolumn{1}{c|}{$-0.011$} &
\multicolumn{1}{c|}{$\sim 0$} &
\multicolumn{1}{c|}{$0.255$} &
\multicolumn{1}{c|}{$\sim 0$} &
\multicolumn{1}{c|}{$\sim 0$} 
\\\hline
\multicolumn{1}{|c||}{$\tilde{D}^{(2)}_{22}$} & 
\multicolumn{1}{c|}{$0.005$} &
\multicolumn{1}{c|}{$\sim 0$} &
\multicolumn{1}{c|}{$-0.01$} & 
\multicolumn{1}{c|}{$\sim 0$} &
\multicolumn{1}{c|}{$-0.006$} & 
\multicolumn{1}{c|}{$0.255$} &
\multicolumn{1}{c|}{$\sim 0$} &
\multicolumn{1}{c|}{$\sim 0$} &
\multicolumn{1}{c|}{$\sim 0$} &
\multicolumn{1}{c|}{$\sim 0$} 
\\\hline
\end{tabular}
\caption{\label{tab1}\protect
  Coefficients of quadratic forms describing the surfaces in Fig.~\ref{fig4}.}
\end{table*}

Introducing non-zero stochastic terms $k_1, k_2\neq 0$ for independent 
stochastic forces, $\Gamma_1$ and $\Gamma_2$, one has diagonal diffusion 
matrices with two eigenvalues, $(k_1 r)^2$ and 
$k_2^2$ at each point $(r,\theta)$ corresponding to the radial and angular 
eigendirections respectively.

Since the eigenvectors do not change when changing the coordinate system
(see Append.~\ref{app:coordinate}), ``mixing'' the radial and angular
coordinates by e.g.\ changing to Cartesian should yield the same 
principal axis. We, therefore, now repeat our analysis for transformed variables   $x=r\cos{\theta}$ and $y=r\sin{\theta}$.

Using It\^o's formulation for the transformation from polar to Cartesian coordinates \cite{vankampen}, the Langevin
system can be defined through the drift and diffusion coefficients in transformed coordinates, $\tilde{h}_{i}(x,y), \tilde{g}_{ij}(x,y)$ (see 
Append.~\ref{app:coordinate}):

\begin{subequations}
\begin{eqnarray}
\tilde{h}_{i}&=& (h_{1}/r-g_{22}^2)Y_i+(-1)^{i}h_{2}(Y_1 \delta_{2i}+Y_2 \delta_{1i}) \label{DriftPolarTransf1}\\
\tilde{g}_{ij}&=& \frac{g_{1j}}{r}(Y_1 \delta_{i1}+Y_2 \delta_{i2})-g_{2j}(Y_1 \delta_{2i}+Y_2 \delta_{1i}) \label{DriftPolarTransf2}
\end{eqnarray}
\label{DriftPolarTransf}
\end{subequations} 
with $i,j=1,2$ and $(Y_1,Y_2)=(x,y)$.
Thus, from $\mathbf{G}$ in the polar coordinate system,  
Eq.~(\ref{DriftPolarTransf2}) yields $\mathbf{\tilde{G}}$
for which the diffusion matrix
$\mathbf{\tilde{D}}^{(2)}=\mathbf{\tilde{G}\tilde{G}}^T$ reads 
\begin{equation}
\mathbf{\tilde{D}}^{(2)}
=
\begin{bmatrix}
 k_{2}^2y^2+k_{1}^2 x^2 & -k_{2}^2xy+k_{1}^2 xy  \\
-k_{2}^2xy+k_{1}^2 xy   &  k_{2}^2x^2+k_{1}^2y^2   
\end{bmatrix}\quad.
\label{CartesD}
\end{equation}
The eigenvalues and corresponding eigenvectors are  
$\tilde{\lambda}_{1}=k_{1}^2(x^2+y^2)$ with 
$\mathbf{\tilde{v}_{1}}=(x/\sqrt{x^2+y^2}, y/\sqrt{x^2+y^2})$ and 
$\tilde{\lambda}_{2}=k_{2}^2(x^2+y^2)$ with
$\mathbf{\tilde{v}_{2}}=(-y/\sqrt{x^2+y^2}, x/\sqrt{x^2+y^2})$.

In contrast to the eigendirections, the eigenvalues depend on the
Jacobian of our transformation, 
since the nonlinear transformation from polar to Cartesian coordinates 
changes the metric.
According to Append.~\ref{app:coordinate} the eigenvalues in polar
coordinates are $\lambda_i=\tilde{\lambda}_i/s_i^2$ for
$i=1,2$ with
$s_1^2=(\frac{\partial x}{\partial r})^2 + (\frac{\partial y}{\partial r})^2=1$
and
$s_2^2=(\frac{\partial x}{\partial \theta})^2 + (\frac{\partial y}{\partial 
\theta})^2=r^2$.
The eigenvalues of our system in Cartesian coordinates are shown 
as solid and dashed lines in Fig.~\ref{fig2} and a full derivation
for the eigenvalues in different coordinate systems is given in
Append.~\ref{app:coordinate} for the general $N$-dimensional case.

Notice that, if we determine the eigenvalues in a Cartesian system, 
the metric is Euclidean, i.e.~the eigenvalues measured in the same 
units in both $x$- and $y$-direction directly characterize the
diffusion in principal directions. 
Nonlinear transformations of coordinates such as the one from Cartesian 
to polar coordinates, however, generally change the metric. 
In such a system the direction of the maximal eigenvalue is not 
necessarily the direction with the highest diffusion.
For example,
in the Hopf-bifurcation in Eqs.~(\ref{HopfSys}), 
the maximal eigenvalue for $r < k_2/k_1$ is in azimuthal
direction, but the maximal diffusion is still in the radial direction.

Looking to $x$ and $y$ as {\it measured} variables, they define
our Cartesian system and therefore $\tilde{\lambda}_1$ is our
maximal eigenvalue and $\tilde{\lambda}_2$ is associated with the
eigendirection showing minimal stochastic fluctuation.

We proceed to verify this result by numerically analyzing
the time series $x$ and $y$ plotted in Fig.~\ref{fig3}, following the
procedure described in the previous section. The procedure is as follows.
First, we compute the time series according to Eq.~(\ref{HopfSys}) in 
coordinates $r$ and $\theta$. Then, we transfer this time series to 
Cartesian coordinates $x,y$. 
In the new coordinates we calculate the 
drift vectors $\tilde{\mathbf{D}}^{(1)}$ and 
diffusion matrices $\tilde{\mathbf{D}}^{(2)}$. 
Finally, having matrices $\tilde{\mathbf{D}}^{(2)}$ at each mesh point, one easily 
computes the eigenvalues and eigenvectors, as shown in Fig.~\ref{fig2}. The analytical results given by Eq.~(\ref{CartesD}) are reproduced nicely:
the stochastic contribution has two eigendirections, one in the radial 
direction and another in the angular one; the larger eigenvalue is associated with
the radial direction. 
The deviations observed for the minimum eigenvalue occur due to
differences in the quality of statistics in different regions of phase space (value of $r$):
in particular regions in phase space at intermediate values of $r$  are less frequently realized in the data series than
others, resulting in a reduced accuracy of the estimated eigenvalues in these regions.

Figures \ref{fig4}a-c show the components of the diffusion matrix,
namely $\tilde{D}^{(2)}_{11}$, $\tilde{D}^{(2)}_{12}$ and $\tilde{D}^{(2)}_{22}$ 
as a function of both $x$ and $y$.
Notice that the  diffusion matrix is symmetric (see Eq.~(\ref{2D})) and therefore
$\tilde{D}^{(2)}_{21}=\tilde{D}^{(2)}_{12}$.
Once the coefficients are numerically derived, one can access their functional 
behaviour and choose a proper basis of functions to fit the data. 
In the case of Fig.~\ref{fig4}, 
the surfaces obtained numerically are  fitted to a full 
quadratic polynomial
in $x$ and $y$ through a least square procedure.
The coefficients of these polynomial fits are also shown in Fig.\ref{fig4}a-c.
The results of Fig.~\ref{fig4}a-c comply with the analytical expression in
Eq.~(\ref{CartesD}).
A similar analysis is done for the drift vector, i.e.~for the function 
$\tilde{D}^{(1)}_1$ and $\tilde{D}^{(1)}_2$, shown in Figs.~\ref{fig4}d-e.
Table \ref{tab1} lists the values obtained for the coefficients of the 
surface fits in Fig.~\ref{fig4}.

\begin{figure*}[htb]
\begin{center}
\includegraphics*[width=\textwidth,angle=0]{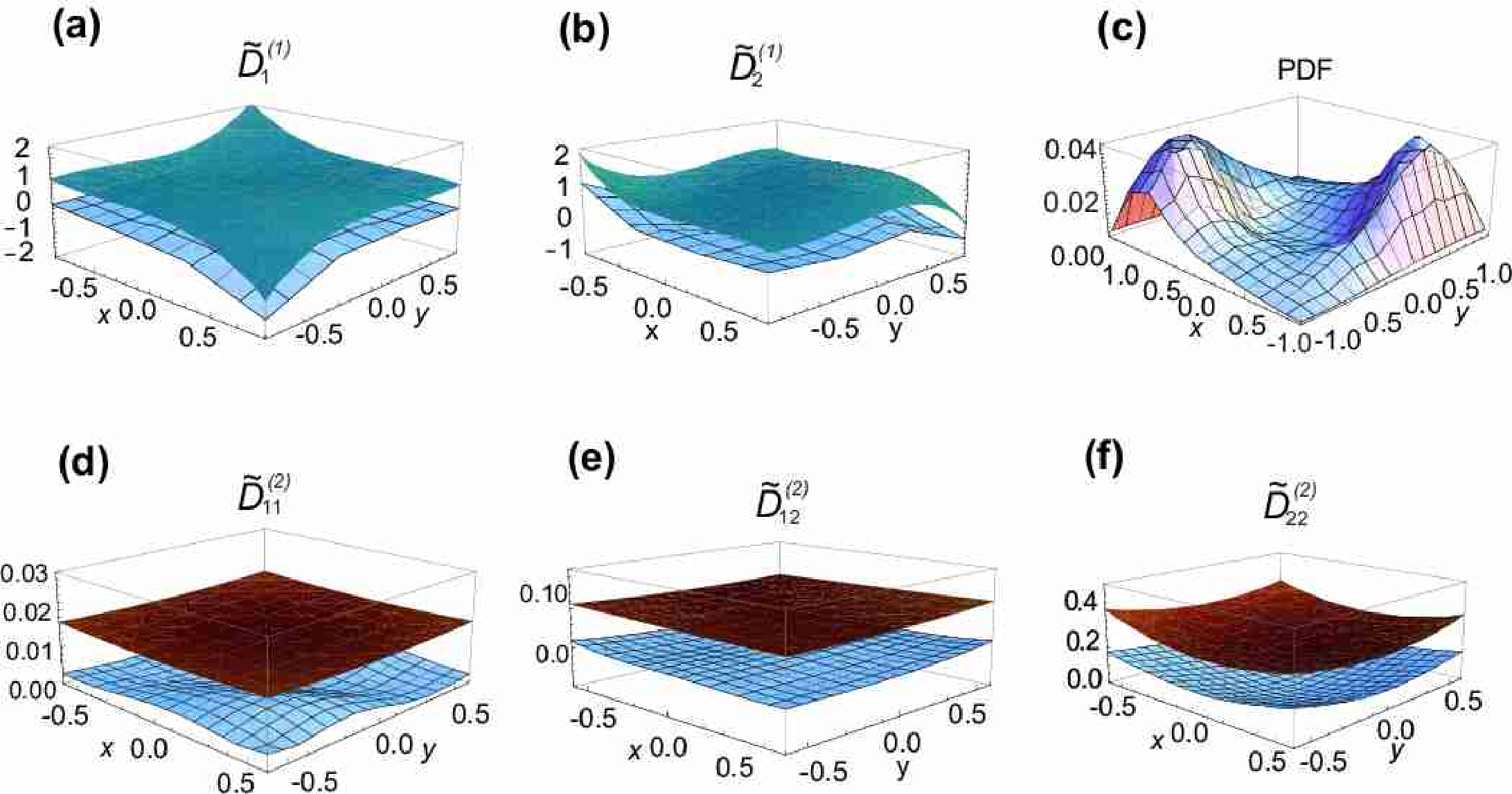}
\end{center}
\caption{\protect
         (Color online)
         Hopf-system with strong stochastic contribution along the 
         $y$-direction.
         Drift functions are
         {\bf (a)} $\tilde{D}_{1}^{(1)}(x,y)$ and
         {\bf (b)} $\tilde{D}_{2}^{(1)}(x,y)$.
         {\bf (c)} Probability density function of $(x,y)$
         shows the most visited regions in phase space.
         Diffusion functions are
         {\bf (d)} $\tilde{D}_{11}^{(2)}(x,y)$,
         {\bf (e)} $\tilde{D}_{12}^{(2)}(x,y)=D_{21}^{(2)}(x,y)$ and
         {\bf (f)} $\tilde{D}_{22}^{(2)}(x,y)$.
         Results were obtained by analyzing the time series 
         $x=r\cos{\theta}$ and $y=r\sin{\theta}$ according
         to Eq.~(\ref{HopfSys2_pol}) with the same upward shifting for
         clarity as in Fig.~\ref{fig4}.
         The probability density function in (c) explains the 
         deviations observed for the diffusion functions (see text).}
\label{fig5}
\end{figure*}
\begin{figure*}[htb]
\begin{center}
\includegraphics*[width=\textwidth,angle=0]{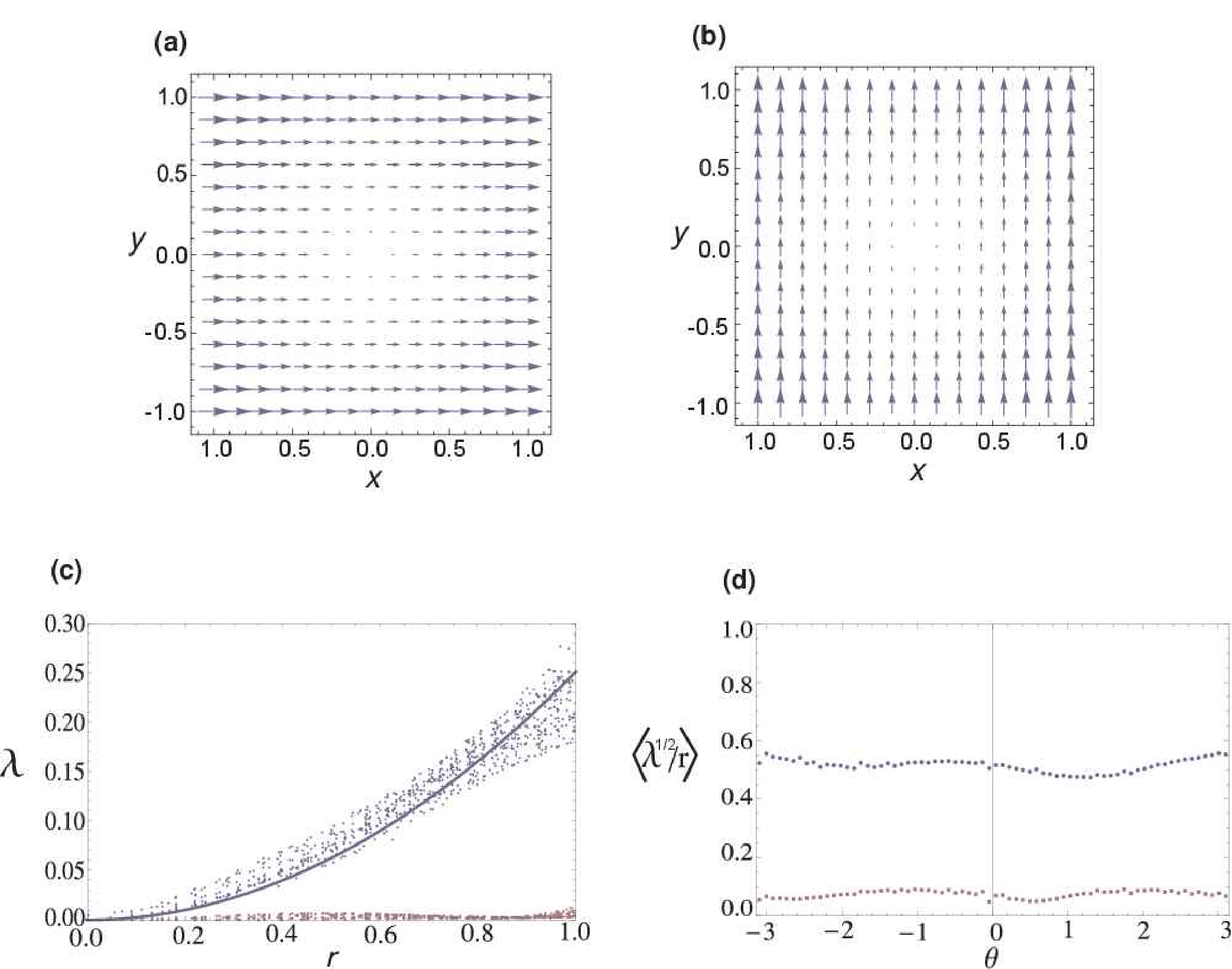}
\end{center}
\caption{\protect
         (Color online)
         Results of the analysis of diffusion matrices plotted in Fig.~\ref{fig5} obtained from a system with an increased level of dynamical noise on the $y$ component (second example). The individual panels depict 
         {\bf (a)} the vector fields of eigenvectors for the minimum and
         {\bf (b)} the maximum eigenvalue, normalized to the respective 
              eigenvalues, and the dependence of the eigenvalues both both on  
         {\bf (c)} $r$ and 
         {\bf (d)} $\theta$.}
\label{fig6}
\end{figure*}

It is worth stressing  that the aforementioned procedure  can likewise  be applied 
to measured multi-dimensional time series in cases where no additional information on the underlying dynamics is available. In this case the  corresponding  fields of eigenvectors indicate 
the directions with largest and smallest stochastic contributions. 
Using this information, a transform of variables to a coordinate system aligned with the direction 
of the smallest stochastic contribution can be applied, which in the present case would 
be the tangential direction. 
In the representation of these new coordinates, the stochastic contribution 
will then be reduced.    

In the previous case, the two eigenvectors for diffusion were 
introduced tangentially and perpendicularly to the limit cycle. 
If the principal axes for the stochastic contribution are not aligned
with the trajectories of the deterministic part of the system, one might suspect  an interplay  
between both stochastic contribution and the drift of the system.

In order to demonstrate the wider applicability of our method,  
 we next consider such a case, with the major axis aligned with 
$y$-direction. We therefore investigate the system
\begin{subequations}
\begin{eqnarray}
\frac{dr}{dt}&=&r(1-r^2) + K_{1} r \cos{\theta} \Gamma_1+ K_{2} r \sin{\theta} \Gamma_2 \\
\frac{d\theta}{dt}&=&(\alpha-r^2) - K_{1} \sin{\theta} \Gamma_1 + K_{2} \cos{\theta} \Gamma_2\quad.
\end{eqnarray}
\label{HopfSys2_pol}
\end{subequations}
In Cartesian coordinates the dynamics are described by
\begin{subequations}
\begin{eqnarray}
\frac{dx}{dt}&=&\tilde{h}_1 + K_{1}r \Gamma_1 \\
\frac{dy}{dt}&=&\tilde{h}_2 + K_{2}r \Gamma_2 .
\end{eqnarray}
\label{HopfSys2_cart}
\end{subequations}
with
\begin{subequations}
\begin{eqnarray}
\tilde{h}_1 &=& x(1-(x^2+y^2))-y(\alpha-(x^2+y^2))\cr
            & & +\frac{x}{x^2+y^2} \left (
                (K_1^2-2K_2^2)y^2-k_2^2x^2
                \right ) \\
\tilde{h}_2 &=& y(1-(x^2+y^2))-x(\alpha-(x^2+y^2))\cr
            & & +\frac{y}{x^2+y^2} \left (
                (K_2^2-2K_1^2)x^2-k_1^2y^2
                \right ) .
\end{eqnarray}
\label{tilde_h}
\end{subequations}
Here we chose $K_1=0.05$ and $K_2=0.5$. 
In other words, the impact of stochastic fluctuations is large in the 
$y$-direction and small along the  $x$-axis. Notice that the determinant
of the diffusion matrix is not preserved, since the Jacobian of
the transformation is not the identity matrix 
(see Append.~\ref{app:coordinate}).

As already done in the previous example, we integrate 
system (\ref{HopfSys2_pol}) numerically  and analyze the resulting data
series of $x$ and $y$ values. 
As can be seen from Fig.~\ref{fig5}, the drift functions $\tilde{h}_1$
and $\tilde{h}_2$ are properly derived (see Figs.~\ref{fig5}a and 
\ref{fig5}b), as well as the terms of the diffusion matrices 
$\tilde{D}^{(2)}$ (Figs.~\ref{fig5}d-f).
The deviations observed for the small $\tilde{D}^{(2)}_{11}$ at particular 
regions of phase space (Fig.~\ref{fig5}d) are due to the lack of accurate 
statistics at those regions (see the PDF in Fig.~\ref{fig5}c).

In both cases presented above the estimates of diffusion
  matrices are in better agreement with the analytical results than
  the estimates of the corresponding drift vectors. This is a general
  drawback of the estimation procedure which comes into play upon
  application to high quality data sets available at high sampling
  frequencies \cite{Kleinhans05Euromech}. Whereas the evaluation of
  the limiting procedure in \eqref{DefCoefKM} converges for the
  diffusion ($k=2$), the error of the drift estimates ($k=1$) diverges
  in the limit $\tau\to 0$ for data sets of finite size due to the
  slow convergence of the law of large numbers implicating small but
  non-zero stochastic contributions of order $\sqrt{\tau}$ to the
  first conditional moments, $\mathbf{M}^{(1)}$\cite{friedrich08}. 
  The drift estimates e.g.\ could be improved by
  advanced estimation techniques applicable at finite time increments
  \cite{Kleinhans05}. Since the method we aim to present, however, is
  based on the properties of the diffusion matrices only this effort
  is not required here.

As can be seen in Fig.~\ref{fig6}a and \ref{fig6}b, the eigendirections
are properly derived as well as the two eigenvalues plotted in 
Fig.~\ref{fig6}c and \ref{fig6}d as a function of $r$ and $\theta$, 
respectively.

With these two examples we first have illustrated that our numerical
approach succeeds in estimating the drift and the diffusion functions
contributions in a two-dimensional stochastic system. A
straightforward computation of the eigenvectors and eigenvalues of the
diffusion matrices then enables us to ascertain a set of principal
stochastic directions.  We would like to note that the method
described here bears some resemblance to the well-known principal
component analysis\cite{pca}. 
In contrast to principal component analysis performed on the distribution 
of the measured data the method we propose, however, focuses on the 
stochastic \emph{fluctuations} in time. The properties of these 
fluctuations in general are dependent on the position $X$ resulting in 
vector fields of the principal axes of the stochastic contributions to 
the dynamics of the system of consideration.
The two methods seem however related and should provide complementary 
insights when applied to specific sets of empirical data. 
Such matters are beyond the scope of this paper and will be addressed 
elsewhere.

We also emphasize
that the minimal eigenvalue of the diffusion matrix not
necessarily points towards the direction the systems as a whole can be
expected to evolve to. It rather reflects the direction of minimal
stochastic forcing only. For prediction of the system's evolution in
general time propagators need to be taken into account which involve
both the stochastic (diffusion) and the deterministic (drift) parts of
the dynamics \cite{friedrich08}.
Nevertheless, being able to ascertain the directions in phase space
where fluctuations are weaker enables one to choose a transformation 
such that part of the new variables have small stochastic terms, 
reducing the number of variables which are affected by stochastic 
forces.
Further, this is also of physical interest to see in which variables 
noise is acting, for example to which variable a thermal bath is 
physically connected, and it may be of technical interest 
to better detect a noise source, for example in an electric 
circuit\cite{friedrich00b}.

\section{Discussion and conclusions}
\label{sec:conclusions}

In this paper we introduce the concept of eigendirections for the stochastic
dynamics in systems of arbitrary dimension. The procedure builds on the modeling of complex systems by means of drift an diffusion functions, that  specify a system of coupled Langevin equations. Estimates for these functions  can be obtained directly from measurements on the systems without prior knowledge on its dynamics following an approach described
in Ref.~\cite{friedrich97,friedrich08}.

In Langevin systems the stochastic forcing is
 composed of independent Gaussian 
$\delta$-correlated stochastic fluctuations. The way how these fluctuations effect the system depends both on the diffusion matrix and on the coordinate system chosen.

Whereas in former publications much attention has been paid to 
imperfections of the Langevin process, like measurement noise or 
correlated noise and finite size sampling\cite{Ragwitz2001,PhysRevLett.89.149401,kleinhans_maximum_2007,PhysRevE.80.031103}
here we introduced
a method which allows us to determine the eigendirections along which each 
stochastic force acts. Each direction of forcing is defined through the eigenvector 
and the corresponding eigenvalue accounting for its amplitude.
The set of eigenvectors does not depend on the choice of the coordinate system
and is therefore  characteristic for the system.
Further, for the particular case where a number $k$ of eigenvalues are 
negligible
in comparison with the others at each mesh point, 
the number of stochastic variables can be
reduced.
Even in cases where the number of stochastic variables cannot be reduced, the eigenvector
associated with
the lowest eigenvalues at each point in phase space indicates the path with minimal stochastic forcing. 
In any case a transform 
to new coordinates in the directions of minimum stochastic forcing can 
be performed, increasing the  relative amplitude of the deterministic components 
and the predictability of the corresponding variable.

The method was successfully applied to a two-dimensional system exhibiting
a Hopf-bifurcation.
For the Hopf-bifurcation the diffusion matrix is better estimated through
our analysis than the corresponding drift vector, contrary to what is
known in many other situations \cite{friedrich08}. 
One note on the dimensionality of each variable in the set of 
variables considered must be stressed: 
If, instead of two position variables, one choses
one position and one velocity or acceleration some caution must be
given to the dimensionality of the corresponding Kramers-Moyal 
coefficients.

For each variable value, Kramers-Moyal coefficients are estimated
from linear least square fits of the conditional moments as functions 
of the time increment within an interval of 
time increments $\Delta t < \tau_{max}$\cite{lind10}.
Consequently, the errors of both first and second Kramers-Moyal 
coefficients associated to the finite-time estimation is of the order
of the correlation coefficient of that least square fit.
Therefore, when computing the eigenvalues of the diffusion matrix, 
there is typically not a significant error propagation and 
consequently the same holds for the eigenvectors. 
Our simulations showed that the relative errors
from finite-time estimation and from the least square fit are 
together typically between $5\%$ and $10\%$.

Nevertheless,
compared to previous methods for minimization of stochasticity  introduced 
in Refs.~\cite{lind05,lind07}, our approach has the advantage of not
requiring a parametrized Ansatz for quantifying the respective stochastic
contributions to the system.
Moreover, it should be applicable  even in the case where the
data sets are contaminated with measurement noise \cite{boettcher06}.

\section*{Acknowledgements}

The authors thank Bernd Lehle for useful discussions.
VVV, FR (SFRH/BPD/65427/2009) and PGL ({\it Ci\^encia 2007})
thank Funda\c{c}\~ao para a Ci\^encia e a Tecnologia (FCT)
for financial support.
All authors thank DAAD and FCT for financial support through 
the bilateral cooperation DREBM/DAAD/03/2009.

\appendix
\section{The diffusion matrix in an arbitrary coordinate system}
\label{app:coordinate}

Consider a transformation of variables 
$\mathbf{X}=\{ X_i \} \to \mathbf{\tilde{X}}=\{ \tilde{X}_i \}$
with $i=1,\dots,N$,
which is given by a two-times continuously differentiable 
deterministic vector function $\f{F}$
\begin{equation}
\f{\tilde{X}}={\f{F}}(\f{X},t)\quad.
\label{transf}
\end{equation}
Using It\^{o}'s formula \cite{fpeq,gard},
a truncated form of the It\^{o} Taylor expansion, the Langevin equations 
for the new variables take the form
\begin{eqnarray}
d\tilde{X}_i  &=&\frac{\partial{F_i}}{\partial{t}}dt+
        \sum_{k=1}^N \frac{\partial{F_i}}{\partial{X_k}}dX_k
       +\cr
      & &\frac{1}{2} \sum_{k=1}^N \sum_{l=1}^N 
         dX_k \frac{\partial{^2F_i}}{\partial{X_k}\partial{X_l}}dX_l +\ldots\quad.
\label{langevin_transf1}
\end{eqnarray}

Eq.~(\ref{Lang2DVect}) can be rewritten to 
\begin{equation}
dX_i = D_i^{(1)}(\f{X})dt +
                  \sum_{j=1}^N g_{ij}(\f{X})\Gamma_j(t)\sqrt{dt}\quad.
\label{langevin2_X}
\end{equation}
\begin{widetext}
Inserting this expression into Eq.~(\ref{langevin_transf1}), retaining all terms in the 
expansion up to order $dt$ while neglecting terms of order
$(dt)^{3/2}$ or higher, and taking advantage of the statistics of the stochastic forcing, Eq.~(\ref{gamma}), one obtains for the third expression on the 
r.h.s.
\begin{equation}
\frac{1}{2} \sum_{k=1}^N \sum_{l=1}^N 
(D_k^{(1)} dt+\sum_{j=1}^N g_{kj}\Gamma_j\sqrt{dt})
\frac{\partial{^2F_i}}{\partial{X_k}\partial{X_l}}
(D_l^{(1)} dt+\sum_{m=1}^N g_{lm}\Gamma_m\sqrt{dt})=
\frac{1}{2} \sum_{k=1}^N \sum_{l=1}^N \sum_{j,m=1}^N 2 g_{kj} g_{lm}
\frac{\partial{^2F_i}}{\partial{X_k}\partial{X_l}}\delta_{jm}dt
\label{hessian}
\end{equation}
Thus, the transformed Langevin equation has the form
\begin{equation}
d\tilde{X}_i  =\frac{\partial{F_i}}{\partial{t}}dt+
       \sum_{k=1}^N \frac{\partial{F_i}}{\partial{X_k}}
       (D_k^{(1)} dt +\sum_{j=1}^N g_{kj}\Gamma_j(t)\sqrt{dt})+ 
        \sum_{k=1}^N\sum_{l=1}^N \sum_{j=1}^N g_{lj}g_{kj}
         \frac{\partial{^2F_i}}{\partial{X_k}\partial{X_l}}dt
        +{\cal O}(dt)^{3/2}
\label{langevin_transf2}
\end{equation}
\end{widetext}

Restricting to stationary processes ($\frac{\partial{F_i}}{\partial{t}}=0$) 
and using the notation
\begin{equation}
\frac{d\tilde{X}_i}{dt} = \tilde{D}_i^{(1)}(\tilde{\f{X}}) +
                  \sum_{j=1}^N \tilde{g}_{ij}(\tilde{\f{X}})\Gamma_j(t)
\label{langeviny}
\end{equation}
the transformed drift function reads
\begin{equation}
\tilde{D}_i^{(1)}= \sum_{k=1}^N \left(D_k^{(1)} \frac{\partial{F_i}}{\partial{X_k}}
             +\sum_{l=1}^N \sum_{j=1}^N g_{lj}g_{kj}
         \frac{\partial{^2F_i}}{\partial{X_k}\partial{X_l}}\right)\quad.
\label{drifty}
\end{equation}
The components of matrix ${\f{{G}}}$ transform as
\begin{equation}
\tilde{g}_{ij}=\sum_{k=1}^N g_{kj}\frac{\partial{F_i}}{\partial{X_k}}
\hspace{0.3cm}
\Leftrightarrow
\hspace{0.3cm} 
\mathbf{\tilde{G}} = \mathbf{J}\mathbf{G}\quad,
\label{gprime-jacobian}
\end{equation}
where $\mathbf{J}$ is the Jacobian of our transformation from coordinates
$X_i$ (vector basis $\mathbf{e}_i$) to $\tilde{X}_i$ 
(vector basis $\tilde{\mathbf{e}}_i$).

From elementary algebra it is known that a vector $\mathbf{v}$ can be 
written in both coordinate systems:
\begin{eqnarray}
\mathbf{v} &=& \sum_j \tilde{v}_j\tilde{\mathbf{e}}_j 
            = \sum_j \tilde{v}_j \sum_k J_{jk} \mathbf{e}_k \cr
           &=& \sum_k \left (
                      \sum_j J_{jk}\tilde{v}_j \right ) \mathbf{e}_k   
            =  \sum_k \left (
                      \sum_j J^T_{kj}\tilde{v}_j \right ) \mathbf{e}_k \cr
            &\equiv& \sum_k v_k \mathbf{e}_k . \label{basistrans} 
\end{eqnarray}

Thus, the matrix $\mathbf{\tilde{U}}$ incorporating the columns of 
eigenvectors $\mathbf{u}_k$ of matrix $\mathbf{\tilde{G}}$, with coordinates
in basis $\tilde{\mathbf{e}}_i$, can be written as
$\mathbf{U} = \mathbf{J^T}\mathbf{\tilde{U}}$
with ${\f U}=[{\f u}_1 \quad {\f u}_2 \quad \dots \quad {\f u}_N] $
fulfilling ${\f U}^{-1}={\f U}^T$.
Thus, from Eq.~(\ref{gprime-jacobian}) one obtains
\begin{eqnarray}
\mathbf{\tilde{U}}^{T}\mathbf{\tilde{G}}\mathbf{\tilde{G}}^T\mathbf{\tilde{U}} &=&
\mathbf{U}^{T}\mathbf{G}\mathbf{G}^T\mathbf{U} .
\label{diagonal}
\end{eqnarray}
So, the eigenvectors of $\mathbf{D}^{(2)}$ and $\mathbf{\tilde{D}}^{(2)}$ 
are the same, apart from the basis in which they are considered,
and in the original (Cartesian) coordinate system $\mathbf{X}$, the 
transformation to principal axes is given by Eq.~(\ref{diagonal}) i.e. 
\begin{equation}  
{\f U}^T {\f D}^{(2)}{\f U} ={\rm diag}[\lambda_1 \quad \lambda_2 \quad \dots \quad \lambda_N] 
\label{orth}
\end{equation}

At regular points the transformation in Eq.~(\ref{transf}) can be 
inverted,
\begin{equation}
{\f X}={\f F}^{-1}(\tilde{\f X})=:{\f f}(\tilde{\f X})\quad.
\label{invtransf}
\end{equation}
By definition ${\f f}(\tilde{\f X})$ is chosen such that the normalized 
eigenvectors are given by
\begin{equation}
{\f u}_k=\frac{1}{s_k}\frac{\partial{\f f}}{\partial{\tilde X_k}}
\quad {\rm with} \quad  s_k=\left \vert
\frac{\partial{\f f}}{\partial{\tilde X_k}} \right \vert .
\label{tangent}
\end{equation}
The Jacobian of ${\f f}(\tilde{\f X})$ can then be written as
\begin{equation}
\tilde{J}_{ik}(\tilde{\f X})=\frac{\partial{f_i}}{\partial{\tilde X_k}} \quad,
\label{inverse_jacobian}
\end{equation}
or in a more convenient form
\begin{equation}
\tilde{\f J}=[{\f u}_1 \quad {\f u}_2 \quad \dots \quad {\f u}_N ]
                 \:{\rm diag}[s_1 \quad s_2\quad \dots \quad s_N ]=:{\f {U s}}.
\label{jacobtilde}
\end{equation}
\begin{figure}[thb]
  \centering
  \includegraphics[width=0.48\textwidth]{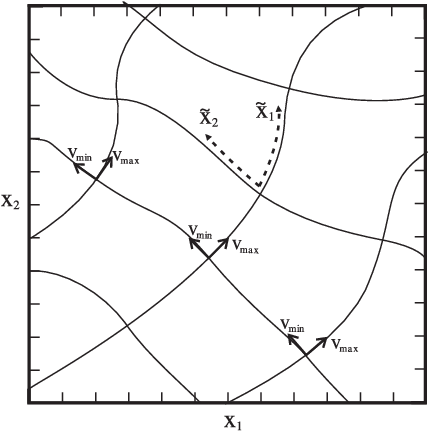}
  \caption{\protect
  Interchanging between coordinate systems. Arrows indicate
  that eigenvectors of the eigenvalues of the diffusion matrix (see
  Eq.~(\ref{Dtilde2})) at each point of the mesh from which drift and
  diffusion coefficients are derived. Here the two-dimensional case
  is illustrated. Extension to arbitrary number of variables is
  straightforward.} 
  \label{fig7}
\end{figure}

The diagonal matrix ${\f s}$ describes the metric of the new system.
The Jacobian of the transformation in Eq.~(\ref{transf}) is the inverse 
relation
\begin{equation}
{\f J}={\f s}^{-1}{\f U}^T \,.
\label{jacob}
\end{equation}
Introducing this relation into Eq.~(\ref{gprime-jacobian}) yields
\begin{equation}
\tilde{\f G}={\f {J G}}={\f s}^{-1}{\f U}^T{\f G}\quad,
\label{Gtilde}
\end{equation}
i.e.~the transformed diffusion matrix reads
\begin{equation}
\tilde{\f D}^{(2)}=\tilde{\f G}\tilde{\f G}^T=
{\f s}^{-1}{\f U}^T{\f G}{\f G}^T{\f U}{\f s}^{-1} \,.
\label{Dtilde}
\end{equation}
Inserting Eqs.~(\ref{orth}) and (\ref{jacobtilde}) finally leads to
the diagonal matrix
\begin{equation}
\tilde{\f D}^{(2)}={\rm diag}\left[\frac{\lambda_1}{s_1^2} \quad 
            \frac{\lambda_2}{s_2^2} \quad \dots \quad 
            \frac{\lambda_N}{s_N^2}\right]  .
\label{Dtilde2}
\end{equation}

In practice, we have no access to the transformation in
Eq.~(\ref{transf}). 
Instead we have a grid (numerical) representation in $\mathbf{X}$ 
coordinate system and search for the transformation in
Eq.~(\ref{transf}) is such that one of the new variables,
say $\tilde{X}_1$, is defined by the field of eigenvectors ${\f u}_1$ 
corresponding to the maximal eigenvalue $\lambda_1$. 
Then, each orthogonal direction is consequently defined
by one of the other (orthogonal) eigenvectors ${\f u}_j$ with
$j=1,\dots,N$. Figure \ref{fig7} illustrates this for the 
two-dimensional case.

To obtain the coordinates of each point $P$ in this grid
in the new coordinates $\mathbf{\tilde{X}}$ one considers the transformation
(\ref{transf}) together with the inverted transformation (\ref{invtransf})
and solves the system of PDEs in (\ref{tangent}).
The problem lies in the coupling introduced by the factors $s_k$ and 
the additional complication of finding the correct boundary conditions
which turns the solution of the PDEs into a complicated problem, even 
numerically, although in 2D the $s_k$ can be neglected as we are
only interested in finding the direction $\frac{\partial f_i}{\partial X_k}$.
In most cases, however, it should be possible to guess a suitable transform 
from  visual inspection of the fields of eigenvectors.
From one point to the next one in the mesh, the eigenvectors 
are sorted according to continuity arguments. 
Further, depending on the obtained fields of eigenvectors,
a scaled polar form or hyperbolic coordinates may be considered, 
where the parameters of said transform can then be fitted to the 
vector fields. Such particular cases depend on the
specific data set at hand and will be addressed elsewhere.




\begin{thebibliography}{00}
\bibitem{friedrich97} R.~Friedrich and J.~Peinke,
                      Phys.~Rev.~Lett.~{\bf 78}, 863 (1997).

\bibitem{friedrich08} R.~Friedrich, J.~Peinke and M.R.R.~Tabar,
                      {\it Complexity in the view of stochastic processes}
                      in 
                      {\it Springer Encyclopedia of Complexity and Systems
                      Science} (Springer, Berlin, 2008).

\bibitem{lind05} P.G.~Lind, A.~Mora, J.A.C.~Gallas and M.~Haase,
                    Phys.~Rev.~E 72, 056706 (2005).        

\bibitem{friedrich00} R.~Friedrich, J.~Peinke and Ch.~Renner,
                        Phys.~Rev.~Lett.~{\bf 84}, 5224 (2000).
                    
\bibitem{ghasemi07} F.~Ghasemi, M.~Sahimi, J.~Peinke, R.~Friedrich, 
                      G.R.~Jafari and M.M.R.~Tabar,
                      Phys.~Rev.~E~{\bf 75}, 060102 (2007).

\bibitem{boettcher06} F.~B\"ottcher, J.~Peinke, D.~Kleinhans, 
                      R.~Friedrich, P.G.~Lind, M.~Haase,
                      Phys.~Rev.~Lett.~{\bf 97} 090603 (2006).

\bibitem{lind10} P.G.~Lind, M.~Haase, F.~B\"ottcher, J.~Peinke, D.~Kleinhans 
                 and R.~Friedrich,
                 Phys.~Rev.~E {\bf 81} 041125 (2010).

\bibitem{carvalho2010} J.~Carvalho, F.~Raischel, M.~Haase and P.G.~Lind,
                       J.~Physics.~Conf.~Ser.~{\bf 285} 012007 (2011).
								
\bibitem{Kleinhans05} D.~Kleinhans, R.~Friedrich , and A.~Nawroth,
                      Phys.~Lett A, {\bf 346} 42-46 (2005).

\bibitem{Gottschall08NJP} G.~Gottschall, M.~W\"achter and J.~Peinke,
                          New J.~Phys.~{\bf 10}(8) 083034 (2008).

\bibitem{Lade09} S.J.~Lade,
                 Phys.~Lett.~A {\bf 373} 3705-3709 (2009).

\bibitem{fpeq} H.~Risken,
               {\it The Fokker-Planck Equation},
               (Springer, Heidelberg, 1984).
               
\bibitem{gard} C.~W.~Gardiner,
               {\it Handbook of stochastic Methods},
               (Springer, Germany, 1997).

\bibitem{Lamouroux09}D.~Lamouroux and K.~Lehnertz,
                     Phys.~Lett.~A {\bf 373} 3507-3512 (2009).
              

\bibitem{gradisek_eigenvectors} J.~Gradi{\~{s}}ek, R.~Friedrich, E.~Govekar 
and I.~Grabec,
Meccanica, {\bf 38},  33 (2003).

\bibitem{vanMourik_eigenvectors} A.~M.~van Mourik,  A.~Daffertshofer, 
and P.~J.~Beek,
Biological cybernetics {\bf 94}, 233 (2006).

\bibitem{mariabook} J.~Argyris, G.~Faust, M.~Haase and R.~Friedrich,
                    {\it Die Erforschung des Chaos}
                    (Springer, Berlin, 2010).
                    
\bibitem{vankampen} N.G.~van Kampen,
                    {\it Stochastic Processes in Physics and Chemistry}
                    (North Holland, 1992).

\bibitem{Kleinhans05Euromech} D.~Kleinhans and R.~Friedrich,
                              in 
                           {\it Wind Energy: Proceedings of the Euromech 
                           Colloquium}, Eds.~J.~Peinke, P.~Schaumann, and 
                           S.~Barth
                           (Springer, Berlin, 2007), pp.~129-133.

\bibitem{pca} H. Kantz and T. Schreiber, 
        {\it  Nonlinear time series analysis}
        Cambridge Univ.~Press, Cambridge  (1997) 

\bibitem{friedrich00b} R.~Friedrich, S.~Siegert, J.~Peinke, St.~L\"uck, 
                       M.~Siefert, M.~Lindemann, J.~Raethjen, G.~Deuschl,
                       G.~Pfister,
                       {\it Phys.~Lett.~A} {\bf 271} 217-222 (2000).

\bibitem{Ragwitz2001} M.~Ragwitz and H.~Kantz, 
Phys. Rev. Lett. {\bf 87}, 254501 (2001)

\bibitem{PhysRevLett.89.149401} R.~Friedrich, C.~Renner, M.~Siefert and J.~Peinke,  
Phys. Rev. Lett. {\bf 89}, 149401 (2002).

\bibitem{kleinhans_maximum_2007} D.~Kleinhans and  R.~Friedrich,  
Physics Letters A {\bf 80}, 368 (2007).

\bibitem{PhysRevE.80.031103} C.~Anteneodo and R.~Riera, 
Phys. Rev. E {\bf 80}, 031103 (2009).

\bibitem{lind07} P.G.~Lind, A.~Mora, M.~Haase and J.A.C.~Gallas,
                   Int.~J.~Bif.~Chaos {\bf 17}(10) 3461-3466 (2007).

\bibitem{mourik06} A.M.~van Mourik, A.~Daffertshofer and P.J.~Beek,
                   Phys.~Lett.~A {\bf 351}(1-2) 13 (2006). 

\end{thebibliography}
\end{document}